\documentclass[a4paper,9pt]{article}
\usepackage{graphicx}
\usepackage{multirow}
\usepackage{booktabs}
\usepackage{INTERSPEECH2021}
\usepackage{stfloats}
\usepackage{color}
\usepackage{float}

\title{ Building Bilingual and Code-Switched Voice Conversion with Limited Training Data Using Embedding Consistency Loss }
\name{Yaogen Yang$^{1,2}$, Haozhe Zhang$^1$, Xiaoyi Qin$^{1,3}$, Shanshan Liang$^2$, Huahua Cui$^4$, Mingyang Xu$^4$, 
Ming Li$^{1,3}$\thanks{corresponding author: Ming Li, ming.li369@dukekunshan.edu.cn}
}
\address{
  $^1$ Data Science Research Center, Duke Kunshan University, Kunshan, China\\
  $^2$ School of Electronics and Information Technology, Sun Yat-sen University, Guangzhou, China\\
  $^3$School of Computer Science, Wuhan University, Wuhan, China\\
  $^4$Advanced Computing East China Sub-center, Su Zhou, China
  }
  
\email{ming.li369@dukekunshan.edu.cn}

\begin{document}

\maketitle
\begin{abstract}
Building cross-lingual voice conversion (VC) systems for multiple speakers and multiple languages has been a challenging task for a long time. This paper describes a parallel non-autoregressive network to achieve bilingual and code-switched voice conversion for multiple speakers when there are only mono-lingual corpora for each language. We achieve cross-lingual VC between Mandarin speech with multiple speakers and English speech with multiple speakers by applying bilingual bottleneck features. To boost voice cloning performance, we use an adversarial speaker classifier with a gradient reversal layer to reduce the source speaker's information from the output of encoder. Furthermore, in order to improve speaker similarity between reference speech and converted speech, we adopt an embedding consistency loss between the synthesized speech and its natural reference speech in our network. Experimental results show that our proposed method can achieve high quality converted speech with mean opinion score (MOS) around 4. The conversion system performs well in terms of speaker similarity for both in-set speaker conversion and out-set-of one-shot conversion. %We also investigate the model’s performance with a low resource dataset. 
Audio samples are available online\footnote{https://yyggithub.github.io/cross-lingual.io/}.

\end{abstract}
\noindent\textbf{Index Terms}: cross-lingual, voice conversion, bottleneck features, embedding consistency loss

\section{Introduction}

Voice conversion (VC) is to convert the voice characteristics of a source speaker into a desired target speaker, while keeping the linguistic contents unchanged. VC has been applied to many fields such as speaker transformation \cite{Arslan1999Speaker}, silent speech interfaces \cite{6220854}, text-to-speech system \cite{674423}, etc. According to whether the source speaker and the target speaker speak the same language, VC can be divided into intra-lingual VC and cross-lingual VC. For intra-lingual VC, most of the existing approaches are variational auto-encoder (VAE)-based \cite{2019ACVAE,tobing2019nonparallel} and Generative adversarial network(GAN)-based \cite{kameoka2018starganvc,lee2020manytomany}.

For cross-lingual VC,  since the source speaker and the target speaker speak different languages, the speech utterances are inherently different in content. Thus, this task is nonparallel in nature. To achieve cross-lingual VC, we are supposed to disentangle speaker characteristics and the content of the source-speech data in the source language, and then replace the speaker characteristics with those of the given target speaker regardless of what languages the target speaker speak \cite{zhao2020voice}. Vector quantization (VQ) based method has been used \cite{115676} for cross-lingual VC between Japanese and English. This approach does not sufficiently preserve the speakers' identity, where the feature space of the converted envelope is limited to a discrete set of envelopes. \cite{inproceedings} proposes a GMM-based cross-lingual VC to generate polyglot speech corpus. Phonemes from the source language are accordingly replaced by acoustically closer phonemes from the target language under GMM-based VC.  Recently, Phonetic PosteriorGram (PPG) based cross-lingual VC \cite{Personalized,zhou}, which takes advantage of the linguistic information of speech data, also achieves high performance. PPG obtained from an English speaker-independent automatic speech recognition (ASR) are regarded as a bridge across speakers and language boundaries in \cite{Personalized}. English PPG is extracted by a deep bidirectional long short-term memory based neural network to be mapped to mel-cepstrums extracted from Mandarin speech of target speaker. However, the converted voice  obtained from mono-lingual linguistic features would have an accent problem. 
%During testing, using the same English ASR system, PPG are is extracted from a generated English speech. 
%\cite{zhou} utilizes bilingual PPG with an average modeling approach for cross-lingual VC. 
All methods mentioned above focus on one-to-one cross-lingual VC and use the conventional vocoder WORLD \cite{Morise2016WORLD} to reconstruct the waveform from predicted spectrum, which leads to lower naturalness on speech quality and fair speaker similarity.

Nevertheless, multi-speaker multi-lingual VC plays an important role for data augmentation in cross-lingual text-to-speech (TTS) and speaker verification task, which motivates us to study high quality multi-speaker multi-lingual VC. In this paper, we adopt a FastSpeech-based \cite{2019FastSpeech} model as our non-autoregressive VC network. Bilingual bottleneck features from an speaker-independent ASR model are used as the linguistic representations for input utterances. In addition, we explore the use of an embedding consistency loss in multi-lingual VC to improve the speaker similarity. Results show that our proposed method can generate fluent, high-fidelity, and intelligible speech in both Mandarin and English provided only mono-lingual training data. 

% The main contributions of this work are listed below:

% \begin{itemize}
% \item[1.] Achieving a multi-speaker multi-lingual VC systems in a parallel  non-autoregressive way.
% \item[2.] Introducing an adversarial speaker classifier and feedback constraint loss to improve high quality converted speech speaker similarity for seen and unseen speaker.

% \end{itemize}

This paper is organized as follows. Section \ref{Related works} describes related works in terms of bilingual linguistic features and speaker verification. Our proposed system is presented in Section \ref{Methods}. Section \ref{Experiments} presents our experimental results. The conclusion is given in Section \ref{Conclusions}.

%\footnote{This research is funded in part by the National Natural Science Foundation of China (61773413), Key Research and Development Program of Jiangsu Province (BE2019054), Six talent peaks project in Jiangsu Province (JY-074), Science and Technology Program of Guangzhou City (201903010040,202007030011).}

\begin{figure*}[htbp]
  \centering
  \includegraphics[scale=0.38]{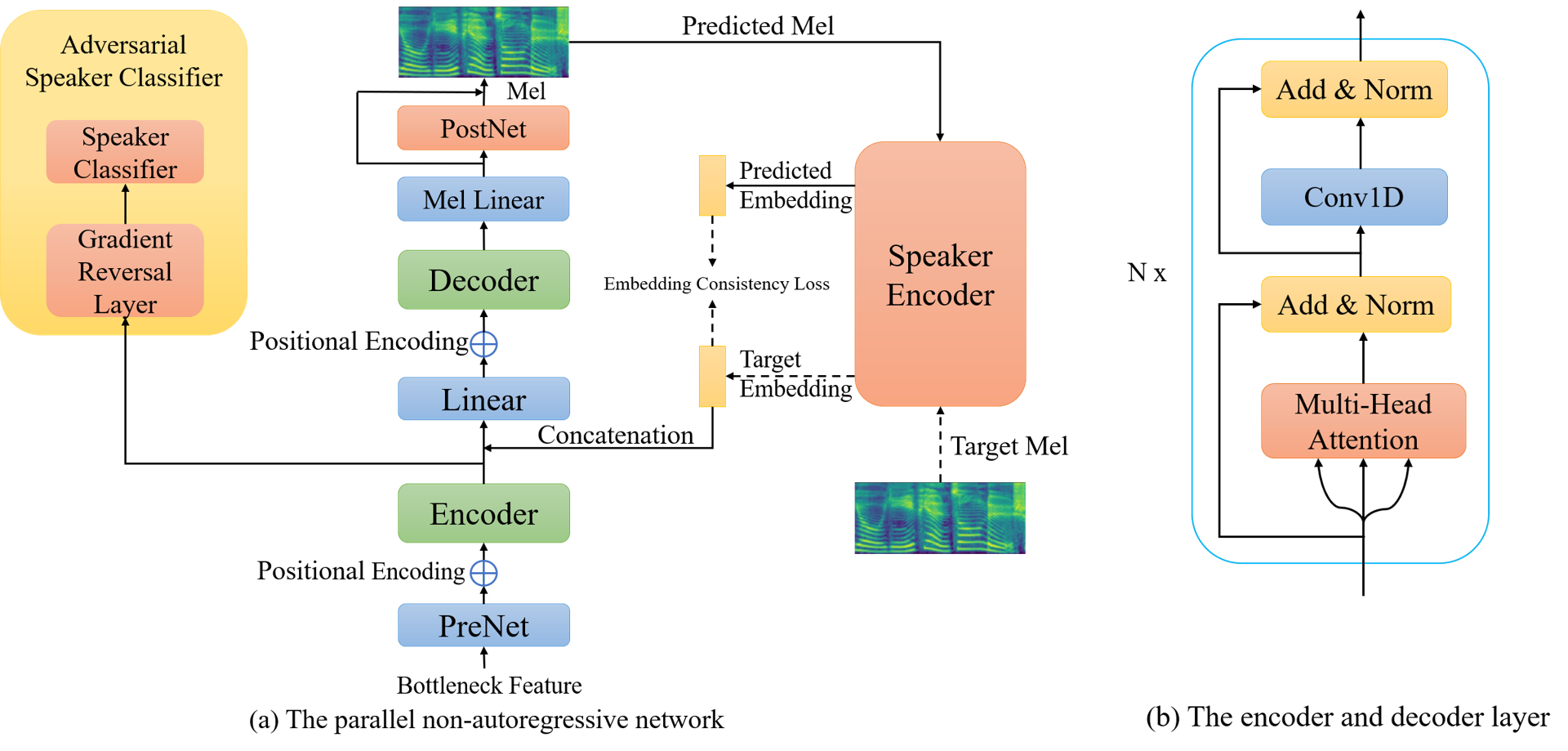}
  \caption{The overall architecture of the proposed method. (a). The parallel non-autoregressive network. (b). The encoder and decoder layer.}
  \label{fig:f1}
\end{figure*}

\section{Related works}
\label{Related works}

\subsection{Bilingual linguistic features}
%%%%% zexin  write  %%%%%
The intermediate linguistic features used in our proposed VC system is the speaker-independent bottleneck feature extracted from a bilingual speech recognition model trained with Kaldi \cite{Povey_ASRU2011}. Typically, speech recognition is trained on audio-text pairs. The recognition process can break down to acoustic feature extraction, phonetic unit prediction and decoding via maximum likelihood estimation with respect to context models like language models. The key module that we borrow from the speech recognition system is the acoustic model that predicts phonetic probabilities from acoustic features. 

In our case, the acoustic model is constructed by time-delayed neural networks (TDNN), where the linear layer before the output layer is designed to be a low-dimensional layer, which is also known as the bottleneck layer \cite{grezl2007probabilistic}. Since the acoustic model is trained to maximize the probability on the true phonetic label for each acoustic frame, the output from the bottleneck layer in a well-trained model should contain precise linguistic information. Thus we can adopt the output of the bottleneck layer as the linguistic feature for voice conversion. 

On the other hand, to extract language-independent feature that works for multilingual scenarios, the speech recognition system is trained on both Mandarin and English data. The phoneme set we adopted for model training is simply the concatenation of phoneme sets from the two languages \cite{vesely2012language}.

%%%%%%%%%%%%%%%%%%%%%%%%%%%

\subsection{Speaker verification}

Speaker recognition is the task of identifying persons from their voices. Recently, deep learning has revolutionized speaker verification. X-vector based \cite{snyder_x-vectors} systems and its variant frameworks \cite{ecapatdnn, cai_exploring_2018} have become the most popular architectures in speaker verification. Normally, the speaker verification contains three main components: a front-end pattern extractor, an encoder layer, and a back-end classifier. The fully connected layer following the pooling layer is adopted as the speaker embedding layer. The output of the fully connected layer is named speaker embedding, which is used as the discriminative fixed-length vector to represent a speaker's identity. Moreover, speaker embedding is also employed in various speech fields, such as target speech separation, multi-speaker speech synthesis, etc. In this paper, we feed the speaker embedding into the cross-lingual voice conversion system to render target speaker's characteristics.

\section{Methods}
\label{Methods}
Non-autoregressive TTS models such as FastSpeech \cite{2019FastSpeech} can synthesize speech significantly faster than autoregressive models with comparable quality. %However, the FastSpeech model relies on an autoregressive teacher model for duration prediction and knowledge distillation to solve the problem of unequal length of phoneme input and acoustic feature output during of the training.
We adopt the FastSpeech \cite{2019FastSpeech} model as our multi-speaker multi-lingual VC network. The framework is shown in Figure \ref{fig:f1}. We remove the length regulator module since the input sequence and the output sequence can share the same length in the VC task. The network is composed of a speaker encoder, encoder-decoder structure with multi-head self-attention mechanism and an adversarial speaker classifier, followed by a vocoder to synthesize the time-domain waveform.

\subsection{Encoder-decoder structure}
FastSpeech is first proposed for converting text embedding sequence to acoustic features, while VC is to convert linguistic features to acoustic features.  %In our work, bottleneck features are extracted by speaker-independent automatic ASR. 
We replace the character-embedding layer with a PreNet, which contains two fully connected hidden layers each with 256 hidden units.
% We also apply dropout layers with dropout rate of 0.5. 
Triangular positional encoding \cite{vaswani2017attention} is added to the output from previous layers to form the input of the encoder and decoder. The encoder contains a stack of N = 4 identical blocks. Each block has two a multi-head self-attention mechanism, followed by two 1D convolutional layer. Residual connections and layer normalization are applied in each convolutional layer. To perform multi-speaker VC, we condition the decoder with a speaker embedding. The speaker embedding concatenated with the encoder output is then fed into a linear layer. The decoder has the same feed-forward network structure as the encoder, which significantly speeds up the training and inference process. A PostNet module consisting of  5-layer convolution is added to take the predicted feature as input to obtain the residual parameters, which improves the overall reconstruction quality. 

%Given a non-parallel multi-speaker multi-lingual corpus, let $x$ be the bottleneck features segment.

% \begin{table}[t]
% \centering
% \caption{The ASR performance of bottleneck extractor}
% \label{tblasr}
% \begin{tabular}{cccc}
% \hline
% \multicolumn{3}{c}{Test Set}      & WER/CER    \\ \hline
% \multicolumn{1}{c|}{\multirow{4}{*}{Librispeech}} & \multirow{2}{*}{dev}  & clean & 3.5\%  \\
% \multicolumn{1}{c|}{}   &       & other & 9\%    \\ \cline{2-4} 
% \multicolumn{1}{c|}{}   & \multirow{2}{*}{test} & clean & 3.9\%  \\
% \multicolumn{1}{c|}{}   &                       & other & 9.32\% \\ \hline
% \multicolumn{1}{c|}{\multirow{2}{*}{AIShell2}}   & \multicolumn{2}{c}{dev}  & 4.29\% \\
% \multicolumn{1}{c|}{}   & \multicolumn{2}{c}{test}     & 4.59\% \\
% \hline
% \end{tabular}
% \end{table}

\subsection{Adversarial speaker classifier }
PPG or bottleneck feature has been proven to contain accent and personalized speaker information
in many-to-many VC systems \cite{Ding2020}. We utilize an adversarial speaker classifier to prevent the converted voice from resembling the source speaker. The adversarial module contains a gradient reversal layer that scales the gradient flowing to the encoder by a factor and a single hidden layer. The last layer produces a probability for each speaker. During forward-propagation, it is optimized to reduce the cross-entropy of speaker classification, and during back-propagation it multiplies
the gradient by $-\lambda$.

%The gradients are clipped to stabilize training. It is optimized to reduce the cross-entropy of speaker predictions. The predictions are done separately for each element of the encoders’ outputs. 

\subsection{Embedding consistency loss}
A feedback constraint mechanism is proposed to improve the speaker similarity between the synthesized speech and its natural reference audio for multi-speaker TTS \cite{cai2020speaker}. Simply concatenating speaker embedding cannot transfer enough speaker information learned by the verification system, especially for cross-lingual VC. Thus, we incorporate the speaker verification model in VC training. We use the embedding consistency loss between the ground truth speaker embedding and the one extracted from the predicted Mel-spectrogram as one component of the loss functions for optimizing the VC network. We
use hyper-paramete $\alpha$ to control the weight of embedding loss.
During the training stage, the parameters of the speaker encoder network are frozen. 

%\begin{equation}
%    L=L_{mel\_loss}+L_{adv\_loss}+\alpha L_{cos\_loss}
%\end{equation}

\subsection{MelGAN vocoder}
A neural vocoder  MelGAN \cite{kumar2019melgan} is used to reconstruct the time-domain waveform from predicted Mel-spectrogram, which can synthesize high-fidelity speech with a fast speed. Our MelGAN network was implemented based on the official open-source code on Github\footnote{https://github.com/descriptinc/melgan-neurips}.

\section{Experimental results}
\label{Experiments}
\subsection{ The bottleneck extractor}
The English dataset Librispeech \cite{panayotov2015librispeech} and the Mandarin dataset AISHELL-2 \cite{du2018aishell} are used to train our bilingual speech recognition model. The receipt that is used to train Librispeech in Kaldi is used for model training. The acoustic model, known as the chain model in Kaldi, has 17 TDNN layers, followed by the 256-dim bottleneck layer. The frame's sub-sampling factor is set to 1 so that the output bottleneck features have the same length as the input acoustic features. The phoneme set we applied includes 39 English phonemes and 52 Mandarin phonemes. We use 50ms window-length and 12.5ms frame-shift for MFCC feature extraction. %The performance of our trained ASR system is shown in table 1. 
The performance of English speech recognition is reported by word error rate (WER), while Mandarin is reported by character error rate (CER). %As shown in table \ref{tabasr}, 
The ASR model achieves low recognition error rates on test sets from the two languages. It achieves a WER of $3.9\%$ on Librispeech test set and a CER of $4.59\%$ on AIShell-2 test. Thus the quality of this acoustic model is acceptable for linguistic feature extraction.

% \begin{table}[htbp] 
%  \caption{The ASR performance of bottleneck extractor.} 
%  \label{tabasr}
%   \centering  
%  \begin{tabular}{p{95pt}p{95pt}}    
%  \toprule   
%   Test Set      & WER/CER \\
%  \midrule   
%   Librispeech    & 3.9\% \\
%   AIShell-2     & 4.59\%  \\
%  \bottomrule 
% \end{tabular}  
% \end{table}

\subsection{ Speaker Encoder}
We adopt the VoxCeleb2 \cite{chung_voxceleb2_2018} dataset to pre-train a large-scale speaker verification system. However, our training dataset is from a different domain (cross-lingual and cross-dataset). To obtain  discriminative speaker embeddings in our cross-lingual setting, we adopt a transfer learning strategy to fine-tune the pre-trained model with our voice conversion training dataset. We employ the ECAPA-TDNN \cite{ecapatdnn} model during the training and test stage, while use another ResNet34-SE \cite{cai_exploring_2018} model to evaluate the objective similarity between converted speech and target speech. The AISHELL-2 and VCTK dataset \cite{veaux2016superseded} are served to fine-tune the model. We split 100 speakers from the AISHELL-2 database as the test set to evaluate the performance of the speaker encoder. Therefore, there are totally 1901 speakers with 950134 utterances for fine-tuning. In the speaker encoder experiments, equal error rate (EER) and minimum detection cost function (mDCF) are used to measure the verification performance. The performance of the speaker system is shown in Table 1. Our denotations are: CH for Mandarin dataset, EN for English dataset.
\begin{table}[htbp] 
 \caption{The performance of speaker encoder system.} 
 \label{tab:spk_encoder_performance}
 \centering    
 \begin{tabular}{lcll}    
 \toprule   
  Model  & EER[\%] & $mDCF$ & Acc[\%] \\
 \midrule   
   %ECAPA-TDNN(CH)  & 1.3362 & 0.2034 & 97.335 \\
   ECAPA-TDNN(CH+EN)  & 2.1800 & 0.4016 & 99.439\\
   ResNet34-SE(CH)  & 1.6450 & 0.1552 & 99.725\\
 \bottomrule 
\end{tabular}  
\end{table}

\subsection{Training setup}
Four public available datasets is used in our experiments, including the LJ Speech (LJS) dataset \cite{ljspeech17}, Blizzard Challenge 2020 (BC2020) dataset \cite{zhou2020blizzard}, the VCTK English dataset \cite{veaux2016superseded} and the AISHELL-3 dataset \cite{shi2020aishell3}. LJS contains approximately 24 hours of audio-transcript English pairs recorded by a female English native speaker. BC2020 contains 4365 audio-text Mandarin pairs about 9.5 hours recorded by a male voice. The VCTK English corpus contains 109 speakers with various accents, and we randomly selected 9 speakers for out-set-of speakers one-shot test. We randomly select 174 speakers from the AISHELL-3 database for training and 10 speakers for out-set-of one-shot test. For code-switch cases, a commercial DB-4 dataset from Data Baker is used as source to provide bilingual speech.
%\footnote{https://www.data-baker.com/us.html#/index}
\begin{figure}[htbp]
  \centering
  \includegraphics[scale=0.40]{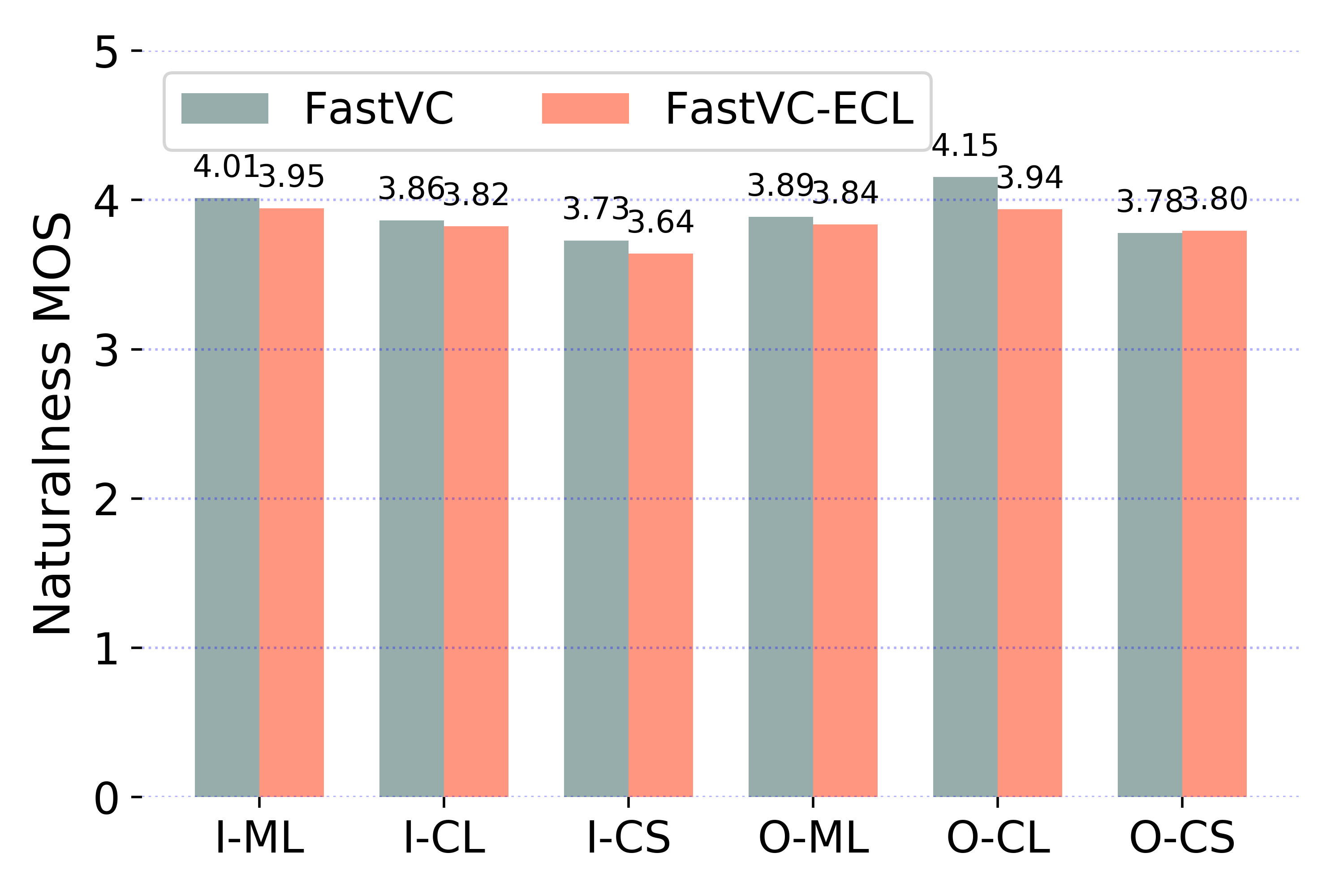}
  \caption{Naturalness MOS result in our experiment. I for in-set test, O for out-set-of one-shot test, ML for mono-lingual VC, CL for cross-lingual VC, and CS for code-switch VC.}
  \label{fig:f2}
\end{figure}

All audios are downsampled to 16 kHz, and all 80-dimensional Mel-spectrogram were extracted every 50ms with windowing of 12.5 ms frame length and 1024-point Fourier transform to obtain the same time sequence length between bottleneck features and Mel-spectrogramin in our training setup. Hyper-parameters $\lambda$, $\alpha$ are set to 1.0, 5.0.

We evaluate the performance by comparing the proposed VC system, a system added embedding consistency loss (ECL) marked as FastVC-ECL, with the multi-speaker VC system without ECL marked as FastVC. We first trained the FastVC model until it can synthesize intelligible speech. Then the FastVC-ECL model is trained from the pre-trained model with the speaker embedding consistency loss.

\subsection{Subjective evaluation}

We conduct Mean Opinion Score (MOS) evaluation on speech naturalness and speaker similarity via listening to converted samples. Participants are asked to rate them on a scale from 1 to 5 with 0.5 increments, and they were advised to ignore the spoken content in cross-lingual similarity test case. 12 native Mandarin speakers, having engaged in speech-related research area for years and being (Mandarin-English) bilingual speakers, were asked to rate the converted speech. We have three types of VC cases for evaluating the performance, which are mono-lingual VC, cross-lingual VC, and code-switch VC, which indicating source speaker speech contains both Mandarin and English content. For each in-set test, each type of VC has 16 randomly picked utterances that are from different speakers. For each out-set-of one-shot test, 8 utterances are randomly picked. 

Figure 2 and 4 shows the MOS scores on speech naturalness and speaker similarity, respectively. %Our notations are: I for in-set test, O for out-set-of one-shot test, ML for mono-lingual VC, CL for cross-lingual VC, and CS for code-switch VC. 
Compared to FastVC system, FastVC-ECL system has maximum 0.1 gap Regardless of whether it is in-set test or one-shot test for speech naturalness.
The FastVC-ECL system obtains naturalness MOS with 3.95 in mono-lingual VC case, but degrades to 3.82 for cross-lingual VC case and degrades to 3.64 for code-switch VC, with similar tendency for one-shot test. For speaker similarity, FastVC-ECL system performs better scores for cross-lingual VC case, but obtain less scores for other cases. The results show that adding embedding consistency loss does not seem to affect the subjective evaluation. 

\begin{figure*}[t]
  \centering
  \includegraphics[scale=0.42]{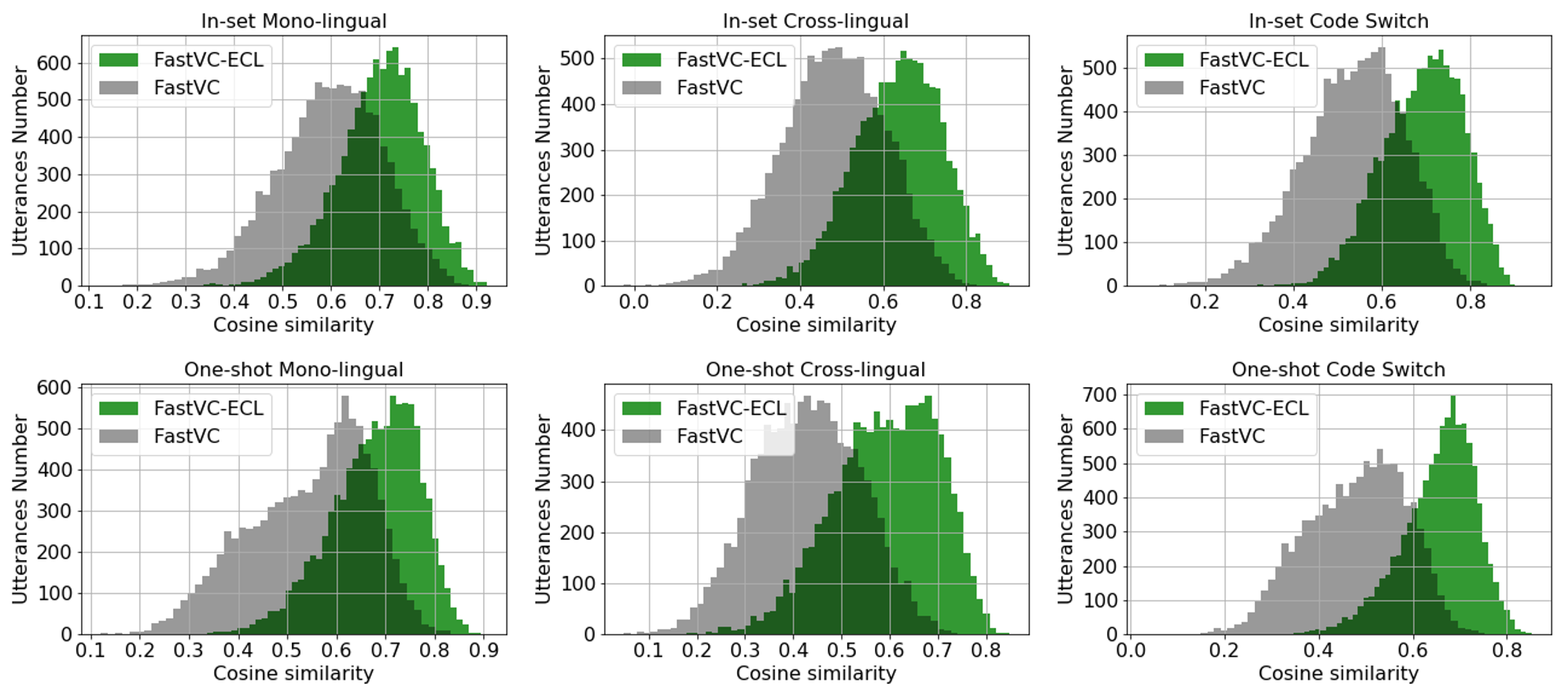}
  \caption{Histogram of cosine similarity score between speaker embeddings from reference speech and converted speech in our experiment.}
  \label{fig:f3}
\end{figure*}

% are shown in Fig.2. These results show that the proposed method significantly outperformed the baseline method in terms of both sound quality and speaker similarity.
% \begin{figure}[H]
%   \centering
%   \includegraphics[scale=0.3]{s-mos.png}
%   \caption{Histogram of cosine similarity score on the experiment.}
%   \label{fig:f4}
% \end{figure}

\subsection{Objective evaluation}

Cosine similarity on speaker embeddings is used for objective assessments in Voice Conversion Challenge 2020 \cite{2020arXiv200903554D}, which extracts speaker embedding vectors from both converted audio and one of the reference audio files using the speaker verification system. Therefore, we computed the cosine similarity of speaker embedding vectors. Similar to subjective evaluation, 10200 utterances are randomly picked from 100 speakers for each type of VC cases in the in-set test. For out-set-of one-shot test, we randomly select 10450 utterances from 19 unseen speakers including mono-lingual VC, cross-lingual VC and code switch VC. 

\cite{cai2020speaker} uses the same speaker verification system adopted in training to evaluate speaker similarity, which can easily lead to overfitting. Thus, we uses ResNet34-SE model which is different from ECAPA-TDNN model used in training to extract speaker embeddings. Figure \ref{fig:f3} shows cosine similarity scores of different testing conditions. Compared to FastVC system, FastVC-ECL system obtains higher scores 
which is close to 0.7 on average in each VC condition. Therefore, the model may be useful for data augmentation and spoofing attack towards to speaker verification.

%Our denotations are: I for in-set test, O for one-shot test, ML for mono-lingual VC, CL for cross-lingual VC, and CS for code-switch VC.
\begin{figure}[htbp]
  \centering
  \includegraphics[scale=0.40]{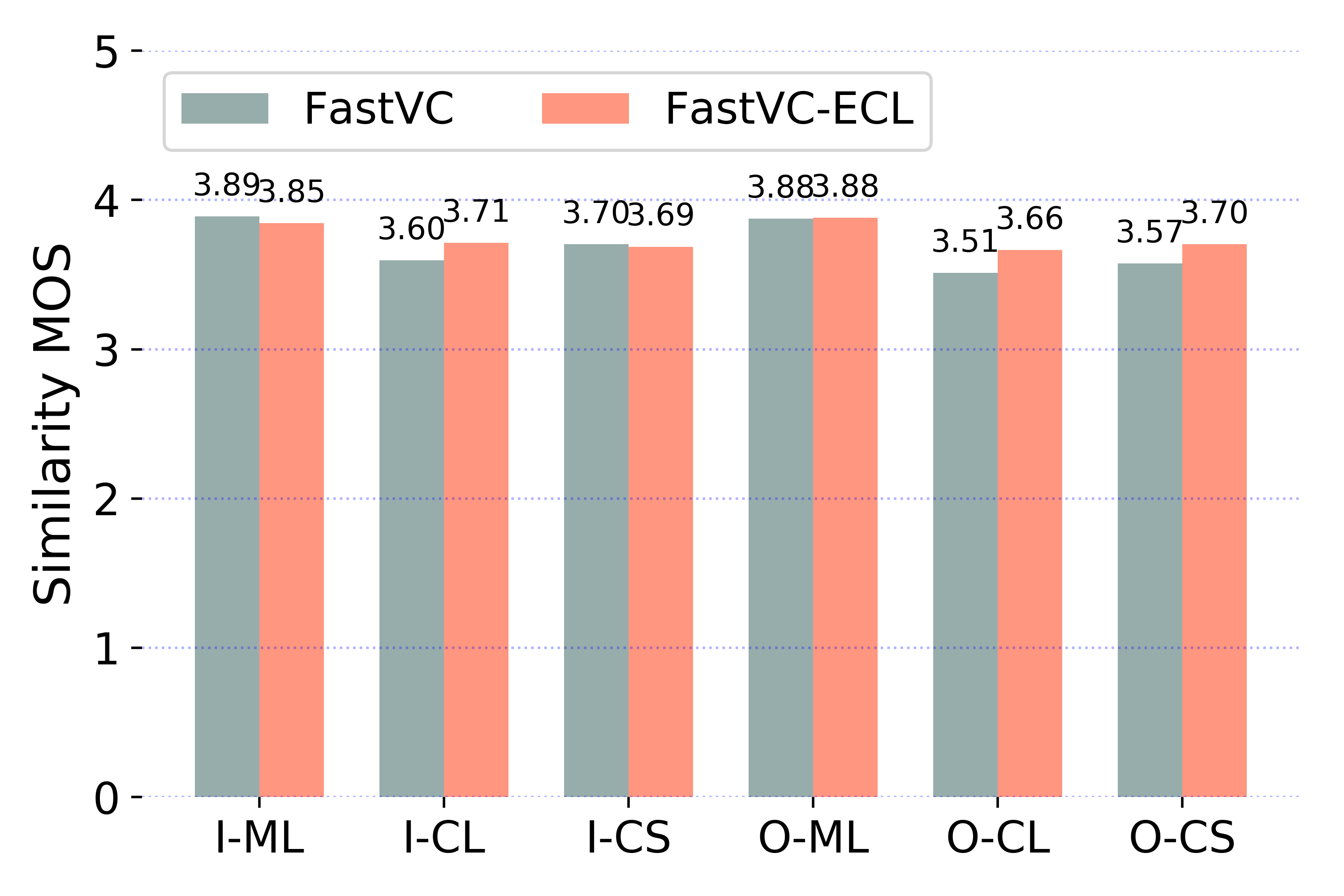}
  \caption{Speaker similarity MOS result in the experiment. I for in-set test, O for out-set-of one-shot test, ML for mono-lingual VC, CL for cross-lingual VC, and CS for code-switch VC.}
  \label{fig:f4}
\end{figure}

\section{Conclusions}
\label{Conclusions}
We present a bilingual multi-speaker VC approach based on a parallel non-autoregressive network. Bilingual bottleneck features are used to address the accent problem in the cross-lingual VC. Our proposed model achieves high quality converted speech with good speaker similarity using mono-lingual training corpus only. We explored the performance of adding embedding consistency loss in VC system, which improves the similarity of speaker embedding for different speaker verification systems. Thus, the converted speech is easier to spoof arbitrary unseen automatic speaker verification system. In the future,  we will explore the proposed VC method in spoofing and countermeasures for automatic speaker verification.
% \section{Acknowledgement}
% This research is funded in part by the National Natural Science Foundation of China (61773413), Key Research and Development Program of Jiangsu Province (BE2019054), Six talent peaks project in Jiangsu Province (JY-074), Science and Technology Program of Guangzhou City (201903010040,202007030011).
\bibliographystyle{IEEEtran}

\bibliography{mybib}

\end{document}